\documentclass[prl,nofootinbib,superscriptaddress,twocolumn]{revtex4}
\usepackage[a4paper,left=1.5cm,right=1.5cm,top=3cm,bottom=3cm]{geometry}

\usepackage{amssymb,amsmath,amsfonts}
\usepackage[dvipsnames]{xcolor}
\usepackage{graphicx}
\usepackage{longtable}
\usepackage{verbatim}
\usepackage{color}

\usepackage{color}
\usepackage{hyperref}
\hypersetup{colorlinks, citecolor=bluscuro, linkcolor=black, urlcolor=bluscuro}
\definecolor{rossos}{cmyk}{0,1,1,0.55}
\definecolor{bluscuro}{rgb}{0.15, 0.2, .85}
\definecolor{bluchiaro}{cmyk}{1,.3,0.,0.1}

\graphicspath{{./Figures/}}
\def\bma#1{\mbox{\boldmath{$#1$}}}

\newcommand{\be}{\begin{equation}}
\newcommand{\ee}{\end{equation}}
\newcommand{\bea}{\begin{eqnarray}}
\newcommand{\eea}{\end{eqnarray}}
\newcommand{\beq}{\begin{equation}}
\newcommand{\eeq}{\end{equation}}
\def\beqa{\begin{eqnarray}}

\def\eeqa{\end{eqnarray}}

\def\lsim{\mathrel{\rlap{\lower4pt\hbox{\hskip0.5pt$\sim$}}
    \raise1pt\hbox{$<$}}}         
\def\gsim{\mathrel{\rlap{\lower4pt\hbox{\hskip0.5pt$\sim$}}
    \raise1pt\hbox{$>$}}}         

\def\pp{{\scriptscriptstyle +}}
\def\mm{{\scriptscriptstyle -}}

\def\Dp{D_{\scriptscriptstyle +}}
\def\Vp{V_{\scriptscriptstyle +}}
\usepackage[normalem]{ulem}

\newcommand{\arXiv}[2]{\href{http://arxiv.org/pdf/#1}{{\tt [#2/#1]}}}
\newcommand{\arXivold}[1]{\href{http://arxiv.org/pdf/#1}{{\tt [#1]}}}

\begin{document}

\title{A Fresh Look at the Calculation 
of Tunneling Actions\\[0.25cm] 
including Gravitational Effects}
\author{J.~R.~Espinosa }
\address{Institut de F\'{\i}sica d'Altes Energies (IFAE), The Barcelona Institute of Science and Technology (BIST),
Campus UAB, 08193 Bellaterra, Barcelona, Spain}
\address{ICREA, Instituci\'o Catalana de Recerca i Estudis Avan\c{c}ats, 
08010 Barcelona, Spain}
\address{Instituto de F\'{\i}sica Te\'orica UAM/CSIC,
Universidad Aut\'onoma de Madrid, 28049, Madrid, Spain}


\begin{abstract}
\noindent
Recently, the calculation of tunneling actions, that control the exponential suppression of the decay of metastable vacua,  has been reformulated as an elementary variational problem in field space.
This paper extends this formalism to include the effect of gravity. 
Considering  tunneling potentials $V_t(\phi)$ that go from the false vacuum $\phi_\pp$ to some $\phi_0$ on the stable basin of the scalar potential $V(\phi)$, the tunneling action is the minimum of the functional $S[V_t]=6 \pi^2m_P^4\int_{\phi_\pp}^{\phi_0}(D+V_t')^2/(V_t^2D)d\phi $, where $D\equiv [(V_t')^2+6(V-V_t)V_t/m_P^2]^{1/2}$,  $V_t'=dV_t/d\phi$ and $m_P$ is the reduced Planck mass. This one-line simple result applies equally to AdS, Minkowski or dS vacua decays and reproduces the Hawking-Moss action in the appropriate cases. This formalism provides new handles for the theoretical understanding of different features of vacuum decay in the presence of gravity.
\end{abstract}

\maketitle


\paragraph{\bf $\bma{ \S\, 1}$ Introduction \label{sec:intro}} 
The calculation of the tunneling action that controls the exponential suppression of the decay of metastable states has been reformulated recently in \cite{me}. The new method offers an alternative to the standard solution by Coleman \cite{Coleman} which is based on the calculation of a tunneling bounce by solving a differential equation in Euclidean space. 

\begin{figure}[t!]
\includegraphics[width=0.9\columnwidth,height=0.6\columnwidth]{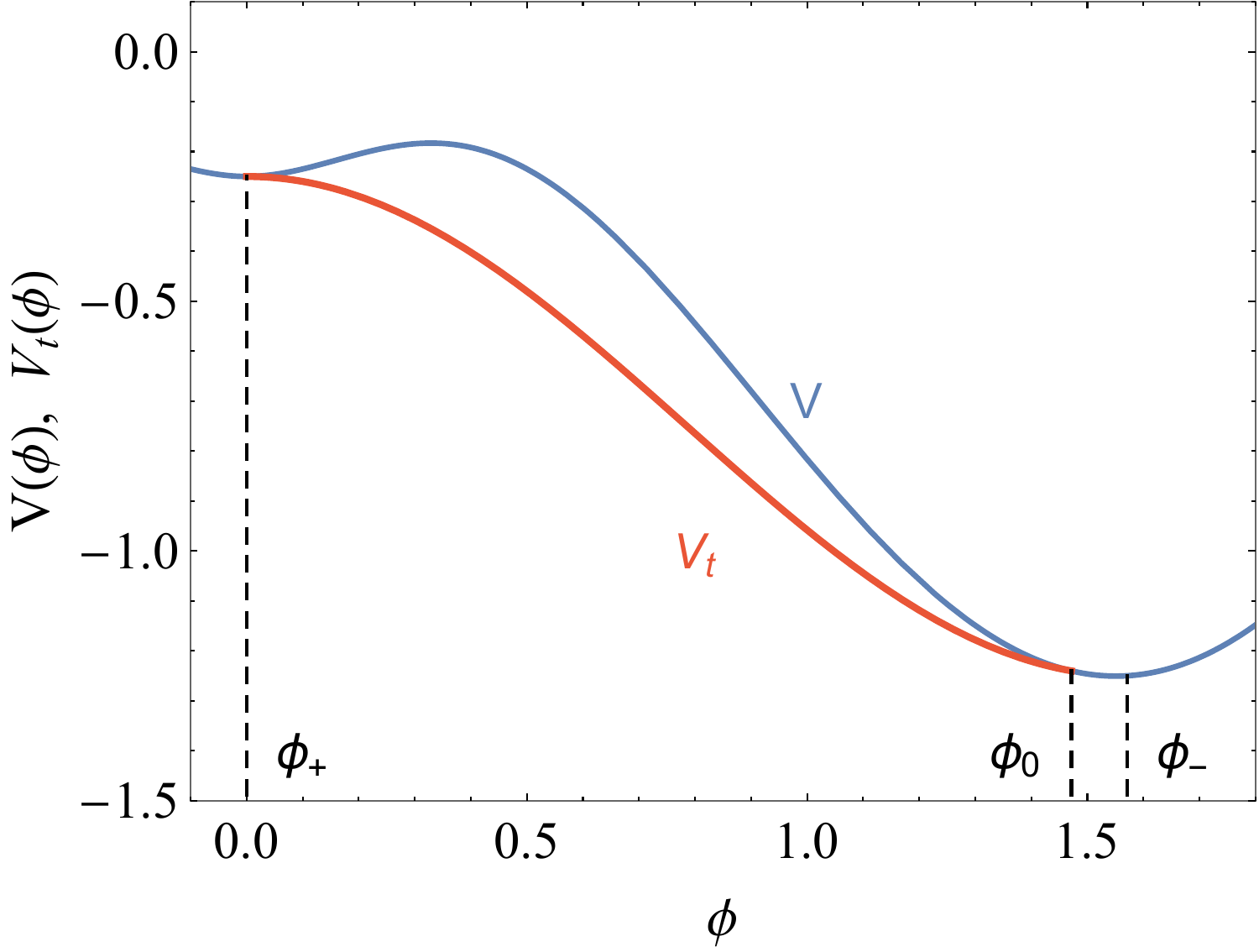}
\includegraphics[width=0.9\columnwidth,height=0.6\columnwidth]{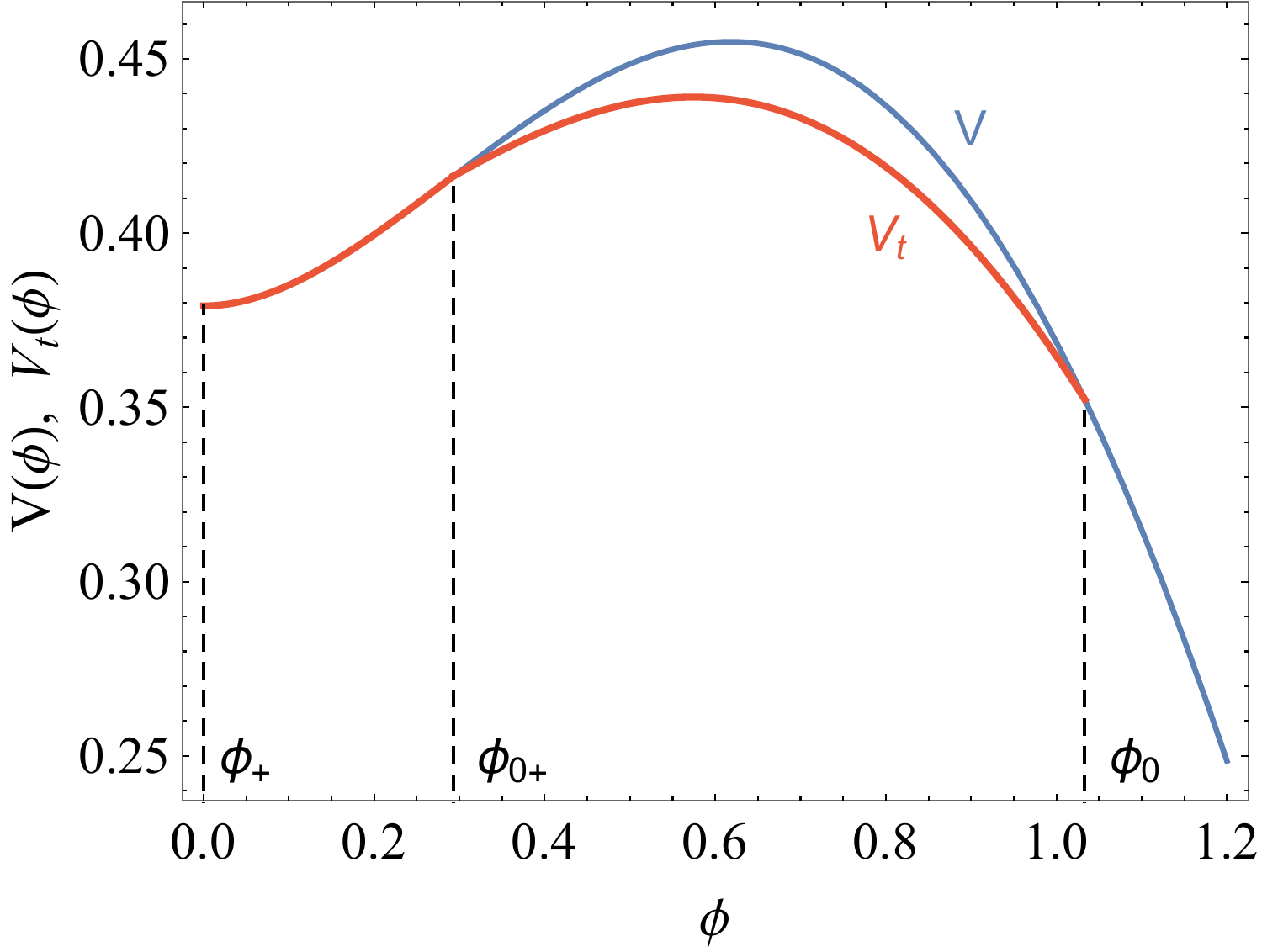}
\caption{\em Tunneling potential $V_t(\phi)$ (red) for decay out of an
AdS vacuum (upper plot) or a dS vacuum (lower plot) in some example potentials $V(\phi)$ (blue).} 
\label{fig:VVt}
\end{figure}

In a nutshell, the new approach works as follows. Consider a potential $V(\phi)$ with a false vacuum at $\phi_\pp$ and a true vacuum at $\phi_\mm>\phi_\pp$, as in the examples shown in Fig.~\ref{fig:VVt}. Take a `tunneling' potential $V_t(\phi)\leq V(\phi)$
that connects the false vacuum with some point $\phi_0$ on the slope beyond the barrier, in the basin of the true vacuum,
with $V_t(\phi_\pp)=V(\phi_\pp)\equiv V_\pp$ and $V_t(\phi_0)=V(\phi_0)\equiv V_0$. 
To such function $V_t(\phi)$ associate the action
\be
S[V_t]\equiv 54\pi^2\int_{\phi_\pp}^{\phi_0}\frac{(V-V_t)^2}{(-V_t')^3}\ d\phi\ ,
\label{SEnograv}
\ee
where $V_t'=dV_t/d\phi$.
Then, under the condition that the action density should satisfy
\be
{\it s}(V_t)\equiv 54\pi^2\frac{(V-V_t)^2}{(-V_t')^3}\geq 0\ ,
\ee
find the $V_t(\phi)$ that minimizes $S[V_t]$.
The minimum action thus found is the tunneling action corresponding to the decay of the false vacuum at $\phi_\pp$. 
For details about the derivation of this simple result, see \cite{me}. 

The new method of calculation has a number of attractive features: it can be considered as a generalization of the thin-wall case for arbitrary potentials; it allows a fast and flexible numerical estimate of $S$, including the case of multifield potentials \cite{EK};
it can be used to generate potentials that admit analytic solutions to the tunneling problem; it can be readily extended to the case of decays by thermal fluctuations, etc., see \cite{me}. Moreover, it is useful to have alternative formulations of important problems as different approaches can offer a better handle in dealing with different issues. Last but not least, the new formulation is extremely direct and simple to state. 

The purpose of this paper is to extend the work in \cite{me} including the effect of gravity, which can be quite relevant for vacuum decay in cosmological settings or in discussions
of the population of vacua in the string landscape, etc. The solution to this problem in the Euclidean bounce formulation dates back to
the work of Coleman and De Luccia in \cite{CdL}.

In the new approach, the inclusion of gravitational effects simply modifies the action density to 
\be
{\it s}(V_t)\equiv \frac{6\pi^2}{\kappa^2}\frac{(D+V_t')^2}{V_t^2 D}\ ,
\label{sEg}
\ee
where $\kappa\equiv 1/m_P^2$, with $m_P=2.435\times 10^{18}$ GeV the reduced Planck mass; $D$ is a generalization of the field derivative of $V_t$ that includes gravitational corrections
\be
D = D(\phi) \equiv \sqrt{(V_t')^2+6\kappa (V-V_t)V_t}\ .
\ee
The problem to solve is as before: find the $V_t$ that minimizes the action $\int_{\phi_\pp}^{\phi_0}{\it s} (V_t)d\phi$. Now the action density (\ref{sEg}) is explicitly positive-definite, but it should be real too, of course, so $V_t$ is constrained to give real $D$.

This remarkably simple formulation applies to the decay of any type of vacua: AdS ($V_\pp<0$), Minkowski ($V_\pp=0$) or dS ($V_\pp>0$) and, in the latter case, it reproduces the Hawking-Moss exponent \cite{HM} in the appropriate cases. Examples of  $V_t$ functions that minimize the tunneling action are shown by the red curves in Fig.~{\ref{fig:VVt}} for an AdS vacuum (upper plot) and a dS one (lower plot).  The qualitatively different behavior of the two cases is apparent (the Minkowski case is similar to the AdS one).

The rest of the paper is organized as follows.
In $\S\, 2$, the standard Euclidean approach of Coleman and De Luccia to the calculation of the semi-classical tunneling exponent
for vacuum decay in the presence of gravity 
is reviewed. In $\S\, 3$, the main result (\ref{sEg})
is obtained, and the corresponding Euler-Lagrange equation for the instanton contribution to the tunneling exponent is derived.
After some comments in $\S\, 4$ on the new action,
  the new method is applied in `reverse gear' to obtain analytically a potential $V(\phi)$ from a simple $V_t(\phi)$ in $\S\, 5$.

After drawing some conclusions, two appendices are devoted to more technical (but important) discussions. Appendix $\S\, A1$ contains the proof that the new formulation agrees with the standard formulation. Appendix $\S\, A2$ derives the thin-wall limit for the tunneling action using the new formulation.

\bigskip
\paragraph{\bf $\bma{\S\, 2}$ Euclidean Action via the Tunneling Bounce \label{sec:bounce}} 

Take a single scalar field $\phi$ in 4 dimensions, with a potential $V(\phi)$ that has a metastable false minimum at $\phi_\pp$ and a deeper minimum at $\phi_\mm$, see Fig.~\ref{fig:VVt}. For simplicity $\phi_\pp=0$ is chosen in many of the plots, without loss of generality. 

The decay of this false vacuum proceeds by fluctuations that nucleate bubbles of the energetically preferred phase which grow and eat out the metastable phase. For sufficiently long-lived states, the vacuum decay rate (per unit volume) $\Gamma/V$ is exponentially suppressed. For the decay by quantum fluctuations  $\Gamma/V=A\ e^{-\Delta S_{E}/\hbar}$, where $\Delta S_{E}$ is the difference between two Euclidean actions, one for the instanton mediating the decay (a bounce/instanton configuration that connects the two phases)  and the other for the false vacuum background.

The bounce configuration results from solving a nonlinear differential equation, an Euler-Lagrange Euclidean equation of motion with appropriate boundary conditions. In the presence of gravity one also has to deal with the cross-talk between the Euclidean space-time metric and the instanton configuration. Assuming that 
the instanton solution that dominates the decay has $O(4)$ symmetry, the most general $O(4)$-symmetric  metric can be written as
\be
ds^2= g_{\mu\nu} dx^\mu dx^\nu = d\xi^2 +\rho(\xi)^2 d\Omega_3^2\ ,
\ee
where $\xi$ is a radial coordinate that measures the radial distance
along lines normal to three spheres of radius of curvature 
$\rho(\xi)$ while $d\Omega_3^2$ is the line element on a unit three-sphere.
The Ricci curvature scalar for this metric is
\be
R=\frac{6}{\rho^2}(1-\rho \ddot\rho-\dot\rho^2)\ ,
\ee
where the dots stand for derivatives with respect to $\xi$.

The instanton action for the decay of the metastable $\phi_\pp$ (false) vacuum is obtained \cite{Coleman} by finding an $O(4)$-symmetric bounce $\phi_b(\xi)$ (or Euclidean bubble) that interpolates between the false vacuum and (the basin of) the true vacuum at $\phi_\mm$. This bounce solves the  Euclidean equation of motion 
\be
\ddot{\phi} +\frac{3\dot\rho}{\rho}\dot{\phi} = V'\ ,
\label{EoM4}
\ee
where a dot (prime) represents a derivative with respect to $\xi$ ($\phi$).
On the other hand, from Einstein's equations it follows that $\rho(\xi)$ satisfies the equation
\be
\dot\rho^2=1+\frac{\kappa \rho^2}{3}\left(\frac12\dot\phi^2-V\right)\ ,
\label{const}
\ee
where $\kappa\equiv 1/m_P^2$.

The system of coupled differential equations (\ref{EoM4}) and (\ref{const}) should be solved with the boundary conditions
\be
\dot\phi_b(0)=0\ ,\quad \rho(0)=0
\ee
at the origin $\xi=0$, with $\phi_0=\phi_b(0)$ an unknown.
For the decay of a dS vacuum, $\rho(\xi)$ returns to zero at some point
\be
\rho(\xi_{\rm max})=0\ ,
\ee
and the bounce is compact.
Instead, for AdS and Minkowski vacuum decay, $\rho(\xi)$ keeps growing indefinitely and the bounce is non-compact. In these cases we take $\xi_{\rm max}=\infty$.
The boundary condition on the bounce at that point is
\be
 \dot\phi_b(\xi_{\rm max})=0\ .
\ee 
In the presence of gravity it is not guaranteed that 
$\phi_b(\xi_{\rm max})=\phi_\pp$: for dS vacua $\phi_b(\xi_{\rm max})\equiv \phi_{0\pp}$ can be quite different from $\phi_\pp$ (although it is always on the basin of the false vacuum).

Identifying $\xi$ with time, Eq.~(\ref{EoM4}) corresponds to the classical motion of a particle in the inverted potential $-V(\phi)$ with a velocity and time dependent friction force. The solution can be found by undershooting and overshooting, changing the value of the field at the center of the Euclidean bubble, $\phi_b(\xi=0)\equiv\phi_0$, till the boundary condition at $\xi\rightarrow \infty$ is satisfied. At the same time, Eq.~(\ref{const}) has to be solved,
with different asymptotic behaviors, depending on whether the false vacuum is dS, Minkowski or AdS.  For AdS and Minkowski vacua,  the bounce is non-compact, with $\rho$ extending over the infinite range $\xi=(0,\infty)$. At $\xi\rightarrow \infty$, for Minkowski vacua $\rho(\xi)\sim\xi$ while for AdS, $\rho(\xi)\sim\exp(\xi/\bar\rho)$, with $1/\bar\rho^2=-V_\pp \kappa/3$. For dS vacua, instead, the bounce is compact and $\rho$ extends over a finite range $\xi=(0,\xi_{\rm max})$, with $\rho=0$ at both ends of the interval. For more details on the standard picture, see \cite{CdL,Weinberg}.

The Euclidean action, from which the previous equations for $\phi$
and $\rho$ follow, is
\be
S_E[\phi] =\int d^4 x \sqrt{g}\left[\frac12 g^{\mu\nu}\partial_\mu
\phi\partial_\nu\phi+V(\phi)-\frac{R}{2\kappa}
\right]+S_{\rm GHY}
\ee
where $S_{\rm GHY}$ is the (Euclidean) Gibbons-Hawking-York boundary term \cite{GHY}. Applying this general expression to the Euclidean bounce, one gets
\bea
S_E[\phi_b] &=& 2\pi^2 \int_0^{\xi_{\rm max}} \left\{
\rho^3\left[\frac12 \dot\phi^2+V(\phi)\right] \right.
\nonumber\\
&+&
\left.\frac{3\rho}{\kappa}\left(\rho \ddot\rho + \dot\rho^2-1\right)
\right\}
 d\xi+ S^b_{\rm GHY}\ ,
\label{SE}
\eea
where it is understood that $\xi_{\rm max}=\infty$ for non-compact bounces, and
\be
S^b_{\rm GHY} = \left.-6\pi^2\frac{\rho^2\dot\rho}{\kappa}\right|_{\xi=0}^{\xi=\xi_{\rm max}}\ .
\ee
Integration by parts gets rid of the $\ddot\rho$ term and cancels
out the GHY term. One gets
\be
S_E[\phi_b] = 2\pi^2 \int_0^{\xi_{\rm max}} \left[
\rho^3\left(\frac12 \dot\phi^2+V\right)
-\frac{3\rho}{\kappa}\left(\dot\rho^2+1\right)
\right]
d\xi\ .
\label{SEbp}
\ee
Furthermore, both expressions (\ref{SE}) and (\ref{SEbp}) for the bounce action can be simplified by using the equation of motion (\ref{EoM4}) and the constraint (\ref{const})
to get the two expressions
\bea
S_{E,1}[\phi_b] &=& -2\pi^2\int_0^{\xi_{\rm max}} \rho^3 V d\xi+S^b_{\rm GHY}\ ,
\label{SE1}\\
S_{E,2}[\phi_b] &=& 4\pi^2\int_0^{\xi_{\rm max}} \left(\rho^3 V-\frac{3}{\kappa}\rho\right) d\xi\ .
\label{SE2}
\eea
The tunneling exponent that suppresses vacuum decay is the difference between this action evaluated for the bounce and the action associated to the background (false vacuum) field configuration. 

The background action can be obtained by substituting in the actions above the potential $V$ by its value at the false vacuum, $V=V_\pp$, and the radius $\rho(\xi)$ by the corresponding solutions of
(\ref{const}) in such background
\bea
\rho_{\rm dS}(\xi) &=& \bar\rho\, \sin(\xi/\bar\rho)\ ,\nonumber\\
\rho_{\rm M}(\xi) &=& \xi\ ,\nonumber\\
\rho_{\rm AdS}(\xi) &=& \bar\rho\, \sinh(\xi/\bar\rho)\ ,
\eea
with $\bar\rho \equiv \sqrt{3/(\kappa |V_\pp|)}$
for decays from dS, Minkowski or AdS vacua, as indicated. One can  obtain the background actions $S_{E\pp}$ analytically. In the case of Minkowski or AdS decays,  which occur through a non-compact bounce (with $\xi_{\rm max}=\infty$), the background action $S_{E\pp}$ diverges, but the divergence cancels against a similar divergence coming from the bounce action $S_{E}[\phi_b]$ and one ends up with a finite tunneling exponent 
\be
\Delta S_E\equiv S_{E}[\phi_b]-S_{E\pp}\ .
\ee
It is customary to regulate such actions using a cutoff in the $\xi$ integrals matching both solutions to enforce the cancelation, but it is  more convenient to leave the $S_{E\pp}$ integral unevaluated and rewrite it in terms of an integral over the bounce $\xi$-coordinate, as shown below.

Consider the AdS vacuum decay first. Using form (\ref{SE1}), the background action for the AdS false vacuum ($V_\pp<0$) can be written
as
\be
S_{E\pp}= -2\pi^2\int_0^\infty \rho_{\rm AdS}^3 V_\pp\ d\xi_{\rm AdS}+S^\pp_{\rm GHY}\ .
\ee
One can change the integration variable from $\xi_{\rm AdS}$ to the bounce $\xi$-coordinate by identifying $\rho_{\rm AdS}(\xi_{\rm AdS})=\rho(\xi)$, from which one gets
\be
\frac{d\xi_{\rm AdS}}{d\xi} = \frac{\dot\rho}{\sqrt{1-\kappa V_\pp\rho^2/3}} \ .
\label{dxiAdS}
\ee
Using this to rewrite $S_{E\pp}$, and noting that $S^b_{\rm GHY}-S^\pp_{\rm GHY}=0$ (both are evaluated at the same $\rho$, and the bounce approaches the false vacuum fast enough),  one ends up with the expression
\be
\Delta S_{E,1} = -2\pi^2 \int_0^\infty \rho^3 \left(
V-\frac{V_\pp \dot \rho}{\sqrt{1-\kappa V_\pp\rho^2/3}}\right)d\xi\ .
\label{DSE1}
\ee

If form (\ref{SE2}) of the action is used instead, following the same
procedure one arrives at
\be
\Delta S_{E,2} = 4\pi^2 \int_0^\infty \left[\rho^3 
V-\frac{3\rho}{\kappa}-\frac{(\rho^3 
V_\pp-3\rho/\kappa)\dot \rho}{\sqrt{1-\kappa V_\pp\rho^2/3}}\right] d\xi\ .
\label{DSE2}
\ee
Of course one should get $\Delta S_{E,1}=\Delta S_{E,2}$.
Both tunneling actions (\ref{DSE1}) and (\ref{DSE2}) also reproduce the Minkowski result simply setting $V_\pp=0$, as the procedure  followed goes through with $d\xi_M/d\xi=\dot\rho$ [which is the $V_\pp=0$ limit of (\ref{dxiAdS})].

For the dS case this rewriting of $S_{E\pp}$ is not needed and one simply has, using (\ref{SE1})
\be
\Delta S_E = S_E[\phi_b]+\frac{24\pi^2}{\kappa^2 V_\pp}\ .
\ee

As in the case without gravity, analytical results for these tunneling exponents are generically not possible and one resorts to numerical solutions of the differential equations. One exception is the case when the false vacuum is nearly degenerate with the true one,
case in which an analytical thin-wall expression can be used. The standard derivation is not discussed here but $\S\, A2$ contains a simple and direct derivation of this formula using the alternative
approach presented in this paper.

\bigskip
\paragraph{\bf $\bma{\S\, 3}$ Euclidean Action via a Tunneling Potential \label{sec:Vint}} 

Let us follow the approach of \cite{me} and introduce an auxiliary function, $V_t(\phi)$, the `tunneling potential'. Its connection 
with the standard bounce method is
\be
V_t(\phi)\equiv V(\phi) -\frac12 \dot\phi_b^2\ ,
\label{Vt}
\ee
where it is understood that $\dot\phi_b$ 
is considered as a function of the field $\phi$.
 
The properties of $V_t(\phi)$ are nearly the same as in \cite{me}:
1) obviously $V_t(\phi)\leq V(\phi)$, with $V_t(\phi_0)= V(\phi_0)$ and $V_t(\phi_{0\pp})= V(\phi_{0\pp})$, as $\dot\phi_b(0)=\dot\phi_b(\xi_{\rm max})=0$ at these end points of the bounce; 2)
$V_t(\phi)$ is a monotonic function for the case of non-compact bounces but it is not-monotonic for compact bounces. This difference in behavior follows from the fact that $V_t(\phi)$ is minus the Euclidean energy, which varies with $\xi$ as
\be
\frac{d}{d\xi}\left[\frac12 \dot\phi_b^2-V(\phi_b)\right]=-\frac{3}{\rho}\dot\rho\, \dot\phi_b^2\ .
\ee
For non-compact bounces $\dot\rho>0$ and the Euclidean energy is dissipated by the friction term in (\ref{EoM4}). However, for compact bounces, $\dot\rho$ changes sign at some intermediate value $\xi=\xi_t$ in the interval $(0,\xi_{\rm max})$. For $\xi>\xi_t$, $\dot\rho<0$ and there is anti-friction.
Noting that the bounce is also a monotonic function of $\xi$ (intuitively clear from the `motion in an inverted potential' picture\footnote{Non-monotonic oscillating Coleman-De Luccia bounces exist \cite{osc} but for the purposes of this paper only their  monotonic part starting at $\xi=0$ is relevant.}), there is a one-to-one correspondence between the slope of $\rho$ and that of  $V_t(\phi)$. So, for non-compact bounces (for decay from AdS or Minkowski vacua) $V_t$ is monotonically decreasing ($V_t'<0$) as in the case without gravity.
For compact cases, corresponding to decays from dS space, the slope of $V_t$ is negative on the side closer to the true vacuum but turns positive on the side closer to the false vacuum (corresponding to $\xi>\xi_{t}$). Examples of these different behaviors  of $V_t(\phi)$ are shown in Fig.~\ref{fig:VVt}.

Proceeding as in \cite{me} one can remove any reference to the bounce (and the  4-dimensional Euclidean space in which it lives) in favor of $V_t(\phi)$. From (\ref{Vt})
\be
\dot\phi_b = - \sqrt{2[V(\phi)-V_t(\phi)]}\ ,
\label{dphi}
\ee
where the minus sign, chosen due to $\phi_\pp<\phi_\mm$, would be a plus if one takes $\phi_\pp>\phi_\mm$. Eq.~(\ref{dphi}) can then be used to remove any $\xi$-derivative of $\phi_b$ in terms of $V_t$ (and $V$). The Euclidean
radius $\rho$ can be obtained from (\ref{EoM4}) and (\ref{const})
as
\be
\rho=\frac{3\sqrt{2(V-V_t)}}{D}\ ,
\label{r}
\ee
where the combination 
\be
D = D(\phi) \equiv \sqrt{(V_t')^2+6\kappa (V-V_t)V_t}\ ,
\ee
first appears.
Further derivatives of $\rho$ can be eliminated in the same way, {\it e.g.}
\be
\dot\rho =-\frac{V_t'}{D}\ ,\quad \ddot\rho = -\frac{\kappa}{3}\rho\ (3V-2V_t)\ .
\ee
Notice that the zero in $\dot\rho(\xi_t)$ for compact bounces is related to a zero in $V_t'$ (at some $\phi_t$ away from the end points), in agreement with the previous discussion. Notice that, to keep $D$ real, this requires $V_t(\phi_t)>0$. 

Taking a derivative of (\ref{r}) with respect to $\xi$ one gets the differential equation for $V_t$:
\be
\boxed{\left(4V_t'-3 V' \right)V_t' = 6(V_t-V)[V_t''+\kappa (3V-2V_t)]}\ ,
\label{VtEoM}
\ee
which takes the place of (\ref{EoM4}) in the new formulation of the tunneling problem (at least for Minkowski and AdS vacua): find a $\phi_0$ and a $V_t(\phi)$ that solve (\ref{VtEoM}) with the boundary conditions:
\be
V_t(\phi_\pp)=V(\phi_\pp)\ ,\quad V_t(\phi_0)=V(\phi_0)\ .
\label{BCV}
\ee
Eq.~(\ref{VtEoM}) also leads [assuming $V'(\phi_\pp)=0$] to
\be
V'_t(\phi_\pp)=0\ ,\quad V'_t(\phi_0)=3 V'(\phi_0)/4\ .
\label{BCVp}
\ee
The case of dS vacua, for which the instanton does not reach all the way to $\phi_\pp$, will be discussed below. 

Note that the limit $\kappa\rightarrow 0$ of Eq.~(\ref{VtEoM}) reduces to the differential equation for the case without gravity discussed in \cite{me}, as it should. Notice also that with the new formulation there is no need to keep track of the two functions $\rho(\xi)$ and $\phi(\xi)$ but just one: $V_t(\phi)$. 

For later use, notice that  Eq.~(\ref{VtEoM}) can be rewritten in terms of $D$ as
\be
\frac{d}{d\phi}\log D = \frac{3V'-4V_t'}{6(V-V_t)}\ .
\label{EoMD}
\ee

Once $V_t$ has been found, one still needs to calculate the action associated to it. In the case without gravity Derrick's theorem \cite{Derrick} was used to select one particular form of the bounce action to be transformed to the new language \cite{me}, but there is no recourse to a Derrick's theorem with gravity. However,
one can reverse engineer the problem and find what action density  reproduces Eq.~(\ref{VtEoM}) under variations with respect
to $V_t$. Following that route, one can determine the tunneling action density up to an arbitrary term that is a total derivative of some function that only depends on $V_t$.
Comparing with  the bounce action of the standard approach that function is fixed and one arrives at
\be
\boxed{
S[V_t]=\frac{6 \pi^2}{\kappa^2}\int_{\phi_\pp}^{\phi_0}\frac{(D+V_t')^2}{V_t^2D}d\phi }\ .
\label{newSE}
\ee
The detailed proof of the exact correspondence between this action and the standard one is presented in $\S\, A1$. In that appendix it is shown that, for the decay of Minkowski or AdS vacua, one has 
\be
S[V_t] = \Delta S_E^{\rm old}\equiv S_E[\phi_b]-S_{E\pp}\ ,
\label{equalSE}
\ee
so that the new action encapsulates in a single expression the difference of the bounce and background actions of the standard formulation.

The dS case requires a more detailed discussion as in general 
$\phi_b(\xi_{\rm max})\equiv\phi_{0\pp}\neq \phi_\pp$, so that the 
field range covered by the compact bounce, $(\phi_{0\pp},\phi_0)$ does not correspond to the range $(\phi_\pp,\phi_0)$ in the integral of Eq.~(\ref{newSE}). Remarkably, an equality like (\ref{equalSE}) also holds in the dS case if one extends the definition of $V_t$ outside the instanton field range, taking
$V_t=V$ in the interval $(\phi_\pp,\phi_{0\pp})$. Such extended $V_t$ is shown in Fig.~\ref{fig:VVt}, lower plot.
With that extension, one has
\bea
S[V_t]&=&\frac{24 \pi^2}{\kappa^2}\int_{\phi_{\pp}}^{\phi_{0\pp}}\frac{V_t'}{V_t^2}d\phi 
+\frac{6 \pi^2}{\kappa^2}\int_{\phi_{0\pp}}^{\phi_0}\frac{(D+V_t')^2}{V_t^2D}d\phi \nonumber\\
&=&\frac{24 \pi^2}{\kappa^2}\left(\frac{1}{V_\pp}-
\frac{1}{V_{0\pp}}\right)+\int_{\phi_{0\pp}}^{\phi_0}{\it s}(\phi)d\phi
\ .
\label{SEdS}
\eea
In $\S\, A1$ it is shown that, again, this exactly reproduces the standard
result, so that (\ref{equalSE}) also holds for dS.

Note that $V_t=V$ is not a solution of the instanton equation of motion (\ref{VtEoM}). Nevertheless,
 an important property of the action (\ref{newSE}) is, by construction, that its functional variation with respect to $V_t(\phi)$ returns the equation of motion (\ref{VtEoM}). More explicitly, one gets
\be
\frac{\delta S}{\delta V_t} = -108 \pi^2 \frac{(V-V_t)}{D^5}\  {\rm{EoM}}\ ,
\ee
where ${\rm{EoM}}\equiv (4V_t'-3V')V_t'+6(V-V_t)[V_t''+\kappa(3V-2V_t)]$. From the additional $(V-V_t)$ factor one sees that the extension $V=V_t$ away from the instanton range also extremizes the action.

Another interesting consequence of Eq.~(\ref{SEdS}) is that 
it returns the Hawking-Moss action when the CdL bounce disappears. In that case, $\phi_{0\pp}$ and $\phi_0$
converge to $\phi_{T}$, the field value corresponding to the top of the barrier. In that case the instanton part of the action (\ref{SEdS})  vanishes and one gets, with $V_T\equiv V(\phi_T)$,
\be
S[V_t]=
\frac{24 \pi^2}{\kappa^2}\left(\frac{1}{V_\pp}-\frac{1}{V_T}\right)\equiv
S_{HM}
\ ,
\label{HMmagic}
\ee
precisely the Hawking-Moss result \cite{HM}.

To see this in more quantitative terms, assume that (the instanton part of) $V_t$ is very flat, say $V_t\simeq V_i$ where $V_i$ is a positive constant (with $V_\pp\leq V_i\leq V_T$). Then one can neglect $V_t'$ and pull $V_t$ out of the integral for $S[V_t]$. Using (\ref{SEdS})
\be
S \simeq \frac{24\pi^2}{\kappa^2}\left(\frac{1}{V_\pp}-\frac{1}{V_i}\right)+\frac{6\pi^2\sqrt{3}\ \sigma(V_i)}{(\kappa V_i)^{3/2}}\ ,
\label{dStw}
\ee
with
\be
\sigma(V_i)\equiv \int_{\phi_{i\pp}}^{\phi_{i-}}\sqrt{2(V-V_i)}\,d\phi\ ,
\ee
where $\phi_{i\pp,i-}$ are the two solutions of $V(\phi)=V_i$.
For a more realistic trial solution, $\sigma(V_i)$ would be replaced by a more complicated integral, but the rough approximation
with flat $V_i$ can be used to estimate parametrically the behavior of the tunneling solution.  

The non-instanton piece of (\ref{dStw}) starts at zero for $V_i=V_\pp$ and grows to the Hawking-Moss action for $V_i=V_T$,
the top of the barrier. The instanton piece instead starts at some positive value for $V_i=V_\pp$ [with $\sigma(V_i)$ corresponding then to the usual wall-tension if the thin-wall limit is applicable], and goes to zero
when $V_i=V_T$. For the restricted type of approximate $V_t$ configurations 
considered, but also in general, the action is minimized by the interplay of these two opposite tendencies. 

If $\kappa V_\pp$ grows [keeping the shape of $V(\phi)$ unchanged], the non-instanton piece of (\ref{dStw}) decreases as $1/(\kappa V_\pp)^2$
while the instanton piece decreases at the slower rate $1/(\kappa V_\pp)^{3/2}$. Therefore, for sufficiently large $\kappa V_\pp$ the non-instanton part wins
and the action is minimized by the Hawking-Moss configuration.

In the opposite limit of $\kappa V_\pp\rightarrow 0$ the non-instanton part diverges and the instanton part dominates the tunneling (with $V_i\rightarrow V_\pp$ and $\phi_{0\pp}\rightarrow \phi_\pp$).

When the minimum of the action occurs at some intermediate value of $V_i$, that value could be calculated by solving 
\be
\frac{d S}{dV_i} = 0\ , 
\ee
which leads to 
\be
 4\sqrt{\frac{V_i}{3\kappa}} = \frac32  \sigma(V_i)+\int_{\phi_{i\pp}}^{\phi_{i-}}\frac{V_i \ d\phi}{\sqrt{2(V-V_i)}} .
\ee
For a thermal description of this two-step dS tunneling see \cite{BW}.

\bigskip
\paragraph{\bf $\bma{\S\, 4}$ Comments on $\bma{S}. $ \label{sec:Comments}} 

\begin{figure}[t!]
\begin{center}
\includegraphics[width=0.9\columnwidth,height=0.6\columnwidth]{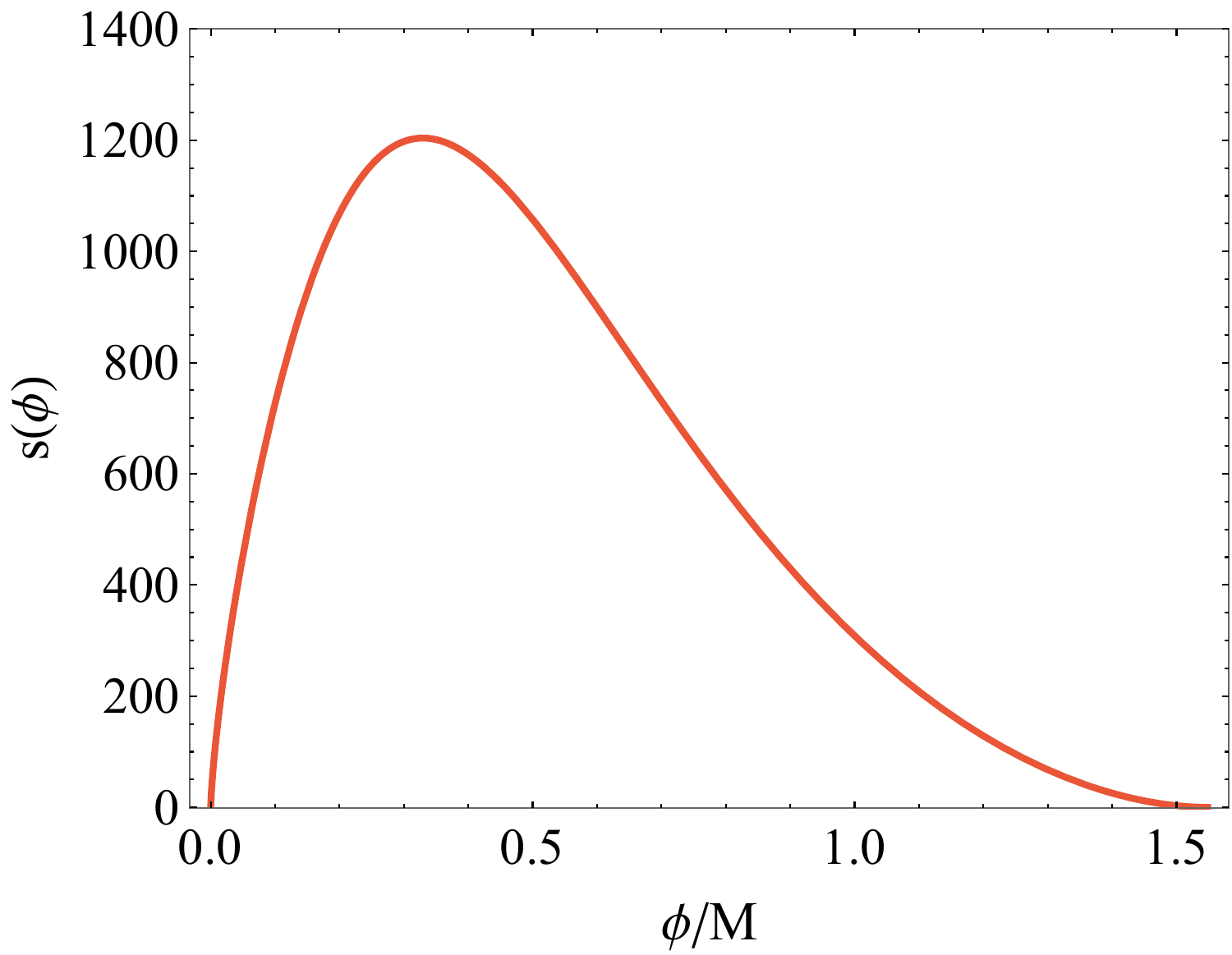}
\includegraphics[width=0.9\columnwidth,height=0.6\columnwidth]{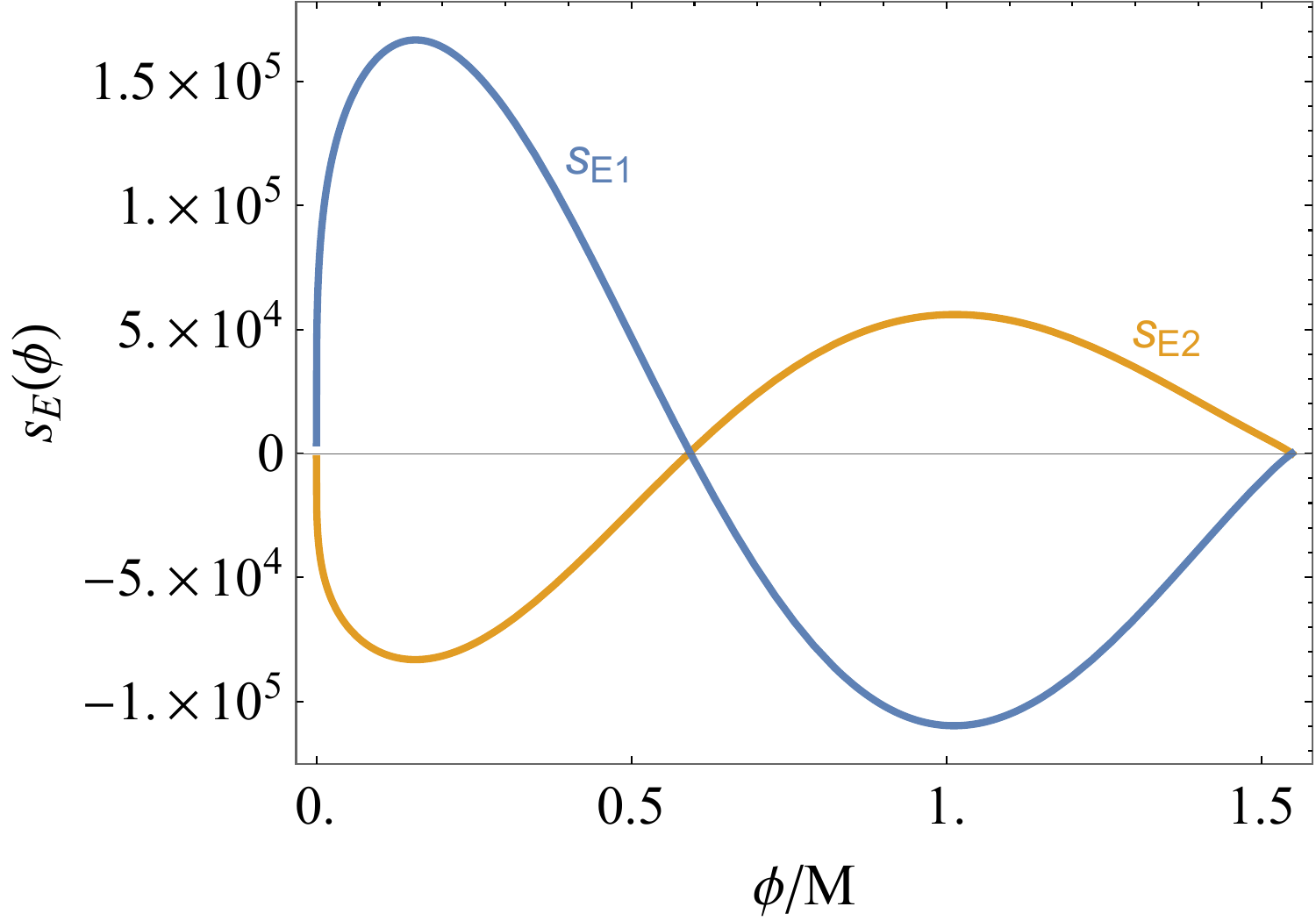}
\end{center}
\caption{\em Comparison of the new action density of Eq.~(\ref{newsE})  (upper plot) with 
standard tunneling action densities 
${\it s}_{E,1,2}$ of Eqs.~(\ref{DSE1}) and (\ref{DSE2}) (lower plot),
for the potential of $\S\, 5$, with $C=100$, $\kappa_0=1$, $V_\pp/\mu^4=-0.5$. }
\label{fig:oldvsnew}
\end{figure}

Some additional comments on the action (\ref{newSE}) are the following.

a) The action density is explicitly positive definite:
\be
{\it s} \equiv \frac{6 \pi^2}{\kappa^2}\frac{(D+V_t')^2}{V_t^2D}\geq 0 \ ,
\label{newsE}
\ee
and should be integrated over a finite field interval.
As a trivial consequence, this expression is better suited for numerical evaluation than the standard ones, which can take both negative and positive values. For illustration, Fig.~\ref{fig:oldvsnew} shows the new action density (red curve, upper plot) and the two versions of the standard action densities corresponding to (\ref{DSE1}) and (\ref{DSE2}), rewritten as functions of $\phi$, (lower plot) for a particular potential (described in the next section). Although the final action is $S\simeq 870$, the standard action densities take very large peak values, which can lead to lower precision of numerical evaluations (that require a cancellation between contributions of opposite sign much larger than the end result).

b) When the gravitational effects are small one can expand the action density in powers of $\kappa$. For the decay from Minkowski or AdS, $V_t'\leq 0$, and the expansion is 
\be
{\it s}
=54\pi^2\frac{(V-V_t)^2}{(-V_t')^3}\left[1-\frac{6\kappa (V-V_t)V_t}{(V_t')^2}\right]+{\cal O}(\kappa^2)\ .
\ee
The $\kappa^0$ term reproduces the action presented in \cite{me}.
Concerning the ${\cal O}(\kappa)$ terms, noting that $V_t\leq 0$ in such decays, one sees that the 
small gravitational effects always make the vacua more stable (higher tunneling action). In assessing this effect there is no need to worry about the ${\cal O}(\kappa)$ effect of gravity on the $V_t$ calculation itself as the zero-gravity action is stationary and such corrections affect the action only at ${\cal O}(\kappa^2)$.

For the decay of dS vacua, there is always a region of field space for which $V_t'\geq 0, V_t\geq 0$. In that region one gets a different expansion:
\bea
{\it s} &=&24\pi^2\frac{V_t'}{\kappa^2 V_t^2}
\\
&+&54\pi^2\frac{(V-V_t)^2}{(V_t')^3}\left[1-\frac{6\kappa (V-V_t)V_t}{(V_t')^2}\right]+{\cal O}(\kappa^2)\ .\nonumber
\eea
As explained before, the $1/\kappa^2$ term plays an important
role in connecting the tunneling action (\ref{newSE}) to the Hawking-Moss action. When the effects of gravity are small, however, $\phi_{0\pp}$ is exponentially close to $\phi_\pp$ and the
divergent contribution from the $1/\kappa^2$ term goes to zero, as we have discussed before.

c) The action density ${\it s}$ would blow up if
either $V_t$ or $D$ go to zero. When $V_t$ crosses zero (as it must happen for a decay from dS to AdS), an expansion in powers of $V_t$ gives
\be
{\it s} = \frac{12\pi^2}{\kappa^2V_t^2}\left[\sqrt{(V_t')^2}+V_t'\right]+{\cal O}(V_t^0)\ .
\ee
The divergence is absent if $V_t'<0$, which is satisfied for dS decays at the point where $V_t$ crosses zero  (remember that 
$V_t'$ starts positive from $\phi_\pp$, reaches zero at some intermediate value where $V_t>0$ and only later becomes negative).

Concerning $D$, which is positive definite, the `equation of motion' for $V_t$, written in the form (\ref{EoMD}), shows that $\log D$ has a finite derivative except where $V=V_t$. So, one does not expect to find $D=0$ at some intermediate value of $\phi$ but $D\rightarrow 0$ will happen at $\phi_\pp$ for AdS or Minkowski decays. In order to get a finite value of $S$ the condition 
\be
\lim_{\phi\rightarrow\phi_\pp} \frac{V_t'}{D}(\phi-\phi_\pp)=0\ ,
\ee
should be satisfied. An explicit case of this behavior is shown in the
discussion of $\S\, A1$.

d) Already from the pioneering work of Coleman and De Luccia \cite{CdL}
it is known that gravitational effects could quench vacuum decay
forbidding decays that would be allowed without gravity (see $\S\, A2$ for a re-derivation of this famous result in the thin-wall limit). In the new formalism, gravitational quenching of vacuum decay can be understood quite simply as the result of the impossibility of finding 
tunneling paths with real $D$ for some potentials. In the Minkowski or AdS cases, the monotonicity of $V_t$ sets a limit on the average values of $V_t'$  and $V_t$ which might not be sufficiently large to compensate the negative value of $6\kappa(V-V_t)V_t$. In integral form, $V_t$ should satisfy the inequality 
\be
\Delta V_{\pp 0}\equiv V_\pp - V_0 \geq \int_{\phi_\pp}^{\phi_0}
\sqrt{6\kappa(V-V_t)(-V_t)}\, d\phi\ ,
\ee
and it is easy to imagine cases in which this is not possible. For instance, large enough $\kappa(V-V_+)$ would lead to gravitational quenching.

e) As in the case without gravity, one expects that it should
be simple to estimate $V_t(\phi)$ for a given potential making educated guesses (as in \cite{me}) to get an accurate approximation to the tunneling action using (\ref{newSE}), although no dedicated study of this application is performed in this paper. The success of such numerical approach rests on the fact that the action (\ref{SEnograv}) is not only an extremal for the $V_t$ that solves the corresponding Euler-Lagrange equation but it is in fact an absolute minimum. In the case with gravity, the action (\ref{newSE}) is, by construction, an extremal for the right $V_t$ that solves (\ref{VtEoM}) (supplemented by an interval with $V_t=V$ in the case of dS vacua). Although we expect that also with gravity present the action (\ref{newSE}) is an absolute minimum at the right $V_t$, we have not been able to prove it in full generality
and leave such proof for future work.

\bigskip
\paragraph{\bf $\bma{\S\, 5}$ Potentials with Exact Tunneling Solutions\label{sec:Exact}} 

Having potentials that allow one to solve the tunneling problem
analytically is quite useful and there is a number of papers in the 
literature that provide such potentials using different methods of attack. In the presence of gravity, a possibly incomplete list of such previous work is \cite{exact}.
 
The new approach to tunneling action calculations based on the tunneling potential is also useful for the purpose of finding such analytical potentials, as was demonstrated in \cite{me}. Here this application is extended to the case with gravity. 

Instead of starting from $V$ and solving for $V_t$, one postulates a given $V_t$ and  integrates (\ref{VtEoM}) to obtain the corresponding $V$. While in the case without gravity a closed form solution for $V$ could be given, this does not seem possible with gravity. Nevertheless, one can still solve for $V$ for particular choices of $V_t$.
 
Take, for instance
\be
V_t = V_\pp -\mu^4 \sin^2\varphi\ ,
\label{Vtan}
\ee
where $V_\pp\leq 0$ (only Minkowski or AdS vacua) and $\varphi\equiv \phi/M$, with $\mu, M$ some mass scales (we take $\mu=M=1$ for numerics) that control the depth and slope of $V_t(\phi)$.
It is possible to integrate (\ref{VtEoM}) obtaining
\be
V = V_t -\frac{\mu^8 s_{2\varphi}^2}{6\kappa_0 V_t}\left\{
1+\frac{\mu^4c_\varphi^{\alpha\kappa_0}s_\varphi^{-(1+\alpha)\kappa_0}}{2[A(\varphi)+C]V_t}\right\}\ ,
\label{Vanalytic}
\ee
where $s_\varphi\equiv \sin\varphi$, $c_\varphi\equiv \cos\varphi$, $\kappa_0\equiv \kappa M^2$, $\alpha\equiv V_\pp/\mu^4-1$. The function $A(\varphi)$ is given in terms of the Appell hypergeometric function of two variables, $F_1(a;b_1,b_2;c;x,y)$, as
\be
A(\varphi)\equiv
\frac{c_\varphi^{2+\alpha\kappa_0}}{\alpha^2(2+\alpha\kappa_0)}
F_1\left(a;b_1,b_2;c;c_\varphi^2;-c_\varphi^2/\alpha\right)\ ,
\ee
with $a=1+\alpha\kappa_0/2$, $b_1=(1+\alpha)\kappa_0/2$, $b_2=2$, and $c=2+\alpha\kappa_0/2$.
Finally, $C$ is an integration constant that can be expressed in terms
of $\phi_0$ solving $V(\phi_0)=V_t(\phi_0)$. 

\begin{figure}[t!]
\begin{center}
\includegraphics[width=0.9\columnwidth,height=0.6\columnwidth]{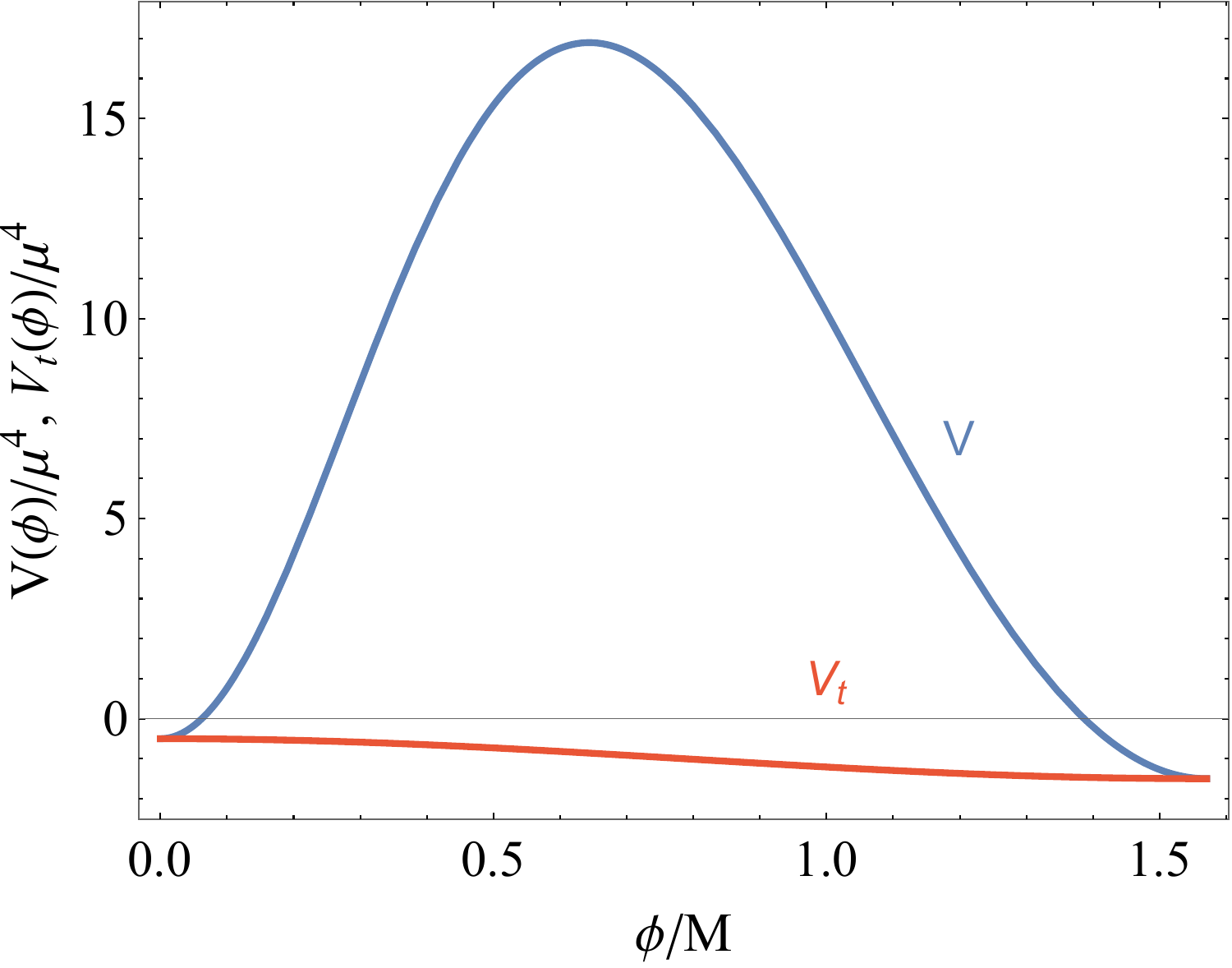}
\end{center}
\caption{\em Potential $V(\phi)$ of Eq.~(\ref{Vanalytic}), with $C=100$, $\kappa_0=0.01$, $V_\pp/\mu^4=-0.5$ (blue) and its corresponding $V_t(\phi)$, from Eq.~(\ref{Vtan}), (red). }
\label{fig:Van}
\end{figure}

An example of this potential is given in the upper plot of Fig.~\ref{fig:VVt}, corresponding to the parameter choice
$C=4$, $\kappa_0=0.8$ and $V_\pp/\mu^4= -0.25$. 
The potential (\ref{Vanalytic}) also has a thin-wall limit for $\kappa_0\ll 1$. As an example, for $C=100$, $\kappa_0=0.01$, $V_\pp/\mu^4=-0.5$ both $V(\phi)$ and $V_t(\phi)$ are plotted in Fig.~\ref{fig:Van}.
 
Rather than a thorough search for the simplest analytical cases, here we just give this example. No doubt other interesting potentials can be found using this method.

\bigskip
\paragraph{\bf $\bma{\S\, 6}$ Conclusions\label{sec:concl}} 

The new formulation of \cite{me} shows how to calculate 
tunneling actions purely in field space, without reference to an
auxiliary Euclidean space in which the standard bounce lives. 
Moreover, it allows one to formulate the problem of obtaining such actions in rather elementary terms, as a quite simple variational  
problem: find a function $V_t$ that minimizes the action functional (\ref{SEnograv}) connecting the metastable and stable phases.

An obvious extension of the work in \cite{me} was to include the effects of gravity, which can be very relevant, and it was not clear whether the simple alternative prescription would also hold in this rather more complicated situation. In this paper it is shown that this is indeed the case, with the problem being reduced to a similar
variational problem, with just a slightly more complicated action density, as given in (\ref{sEg}). It is remarkable that such a simple
prescription encompasses decays from all types of vacua (dS, Minkowski, AdS) in a single universal formula and also reproduces the Hawking-Moss exponent in the appropriate limit. Sidney Coleman would have been pleased with it!

In spite of its very appealing features in terms of simplicity and ease of implementation, this new method cannot be taken as an alternative to completely replace the bounce approach, which is certainly more fundamental. To begin with, the alternative method relies on the $O(4)$ symmetry of the bounce that dominates the decay (an assumption that remains unproven in the case with gravity) and cannot be applied if that symmetry is not present. Moreover, the quantum corrections \cite{ColemanQ} to the semiclassical decay rate have to be computed by considering fluctuations over the bounce and these are explicitly not $O(4)$ symmetric. 

Rather than as a replacement for the bounce method, the new approach is complementary to it and can be most useful not only in numerical applications but also in gaining theoretical insight in particular problems in which such insight is  more difficult to gather from the standard approach. Hopefully it will be fruitful also in this respect. 

\bigskip
\begin{acknowledgments}
\paragraph{Acknowledgments.} I thank Pepe Barb\'on and Jean-Fran\c cois Fortin for interesting discussions and comments. This revised version has benefitted from the use of 
Claude Opus 5.5.
This work has been supported by the ERC
grant 669668 -- NEO-NAT -- ERC-AdG-2014, the Spanish Ministry MINECO under grants  2016-78022-P and
FPA2014-55613-P, the Severo Ochoa excellence program of MINECO (grants SEV-2016-0588 and SEV-2016-0597) and by the Generalitat de Catalunya grant 2014-SGR-1450. 
\end{acknowledgments}

\bigskip
\paragraph{\bf $\bma{\S\, A1.}$ Proof that $\bma{S}\bma{=}\bma{\Delta S_E}$. \label{sec:newold}} 
That the new tunneling action, given in (\ref{newSE}), agrees with the conventional one can be proven by showing that the two action densities differ by an exact differential that vanishes at the boundaries.
Let us write the action functional (\ref{newSE}) as
\be
 S=\int_{\phi_\pp}^{\phi_0}{\it s}\,d\phi\ ,
 \label{action}
\ee
with the (positive definite) action density
\be
{\it s}=
\frac{6 \pi^2}{\kappa^2}\frac{(D+V_t')^2}{V_t^2 D}\ .
\label{senew}
\ee
Consider first the decay from Minkowski or AdS vacua and take as the conventional action Eq.~(\ref{DSE1}). Transforming the integral to a field-space integral using the procedure explained in the main text
one gets
\be
\Delta S_E=\int_{\phi_\pp}^{\phi_0}{\it s}_E\,d\phi\ ,
\ee
with
\be
{\it s}_E= -108\pi^2\frac{(V-V_t)}{D^3}\left(
V+\frac{V_t'\Vp}{\Dp}\right)\ ,
\ee
where 
\be
\Dp \equiv \sqrt{D^2-6\kappa(V-V_t)\Vp }\ .
\ee
Then it is straightforward to check that, for $V_t$ satisfying its equation of motion,
\be
{\it s} - {\it s}_E = \frac{dF}{d\phi}\ , 
\label{senewold}
\ee
with 
\bea
F(\phi)&=&-\frac{12\pi^2}{\kappa^2 V_t}\left[1-\frac{V_t}{\Vp}+\frac{3V_t'}{2D}-\frac{(V_t')^3}{2D^3}
\right.\nonumber\\
&+&\left.
\frac{V_t \Dp}{2V_\pp D}\left(3-\frac{\Dp^2}{D^2}\right)\right]\ .
\eea
To find $F(\phi)$ it was useful to follow the homotopy operator method, as explained {\it e.g.} in \cite{exactd}. In
the Minkowski case (for $V_\pp\rightarrow 0$),
this function has the finite limit:
\be
F_0(\phi)=-\frac{6\pi^2}{\kappa^2 V_t}\left(1+\frac{V_t'}{D}\right)^2\left(2-\frac{V_t'}{D}\right)\ .
\label{F0}
\ee

Integrating 
(\ref{senewold}) in $\phi$ one gets
\be
 S - \Delta S_E  = F(\phi_0) - F(\phi_\pp)\ .
\ee
Noting that $D(\phi_0)=D_\pp(\phi_0)=-V_t'(\phi_0)$, it follows that $F(\phi_0)=0$. To evaluate $F(\phi_\pp)$, notice
that one has $D(\phi_\pp)=D_\pp(\phi_\pp)=0$ 
[assuming that  $V'(\phi_\pp)=0$ at the false vacuum]. To calculate the ratios of these quantities, consider first the AdS case.
Both $V_t'/D$ and $D_+/D$ diverge at $\phi\rightarrow\phi_\pp$
and one needs to know in more detail how $V_t',D, D_\pp$ approach zero. 

Close to $\phi_\pp$, let us approximate the potential by keeping up to its second derivative
\be
V(\phi) = V_\pp + \frac12 m^2 (\phi-\phi_\pp)^2 +\dots
\ee
Solving the equation of motion (\ref{VtEoM}) for the above potential 
leads to the following expansion for the tunneling potential\footnote{This is a Frobenius type of expansion, as expected given the fact that $\phi_\pp$ is a regular singular point of (\ref{VtEoM}).}
\be
V_t(\phi) = V_\pp + \frac12 B (\phi-\phi_\pp)^2 + B_\alpha (\phi-\phi_\pp)^{2+\alpha}+\dots
\ee
with 
\be
B=\frac{3\kappa V_\pp}{2}\left(1+\sqrt{1-\frac{4m^2}{3\kappa V_\pp}}\right)<0\, ,
\label{B}
\ee
and
\be
\alpha = \frac{2\kappa V_\pp}{B}>0\ .
\label{a}
\ee
On the other hand, $B_\alpha$ is a free constant that cannot be determined
by solving (\ref{VtEoM}) around $\phi_\pp$ but is fixed instead by
the boundary condition at $\phi_0$. Such behavior is expected, as $V_t(\phi)$
must depend on the shape of the potential far from $\phi_\pp$. 
From this result it follows that 
\be
V_t',D_\pp \sim (\phi-\phi_\pp)\ , \quad D\sim (\phi-\phi_\pp)^{1+\alpha/2}\ ,
\ee
which shows how the leading terms of $(V_t')^2$ and $6\kappa (V-V_t)V_t$ cancel out in $D$ which is then controlled by the subleading term.

The expansion of $F(\phi_\pp)$ around $\phi_\pp$ gives terms that are clearly zero except for a term proportional to
\be
\frac{(V_t')^3}{D^3}(\phi-\phi_\pp)^2\sim (\phi-\phi_\pp)^{2-3\alpha/2} \ .
\label{ratio} 
\ee
From Eqs.~(\ref{B}) and (\ref{a}), it follows that $\alpha<4/3$ and
so the quantity (\ref{ratio}) goes to zero for $\phi\rightarrow \phi_\pp$ ensuring that $F(\phi_\pp)=0$.

An alternative way of obtaining the same result  is to transform $F(\phi)$ back to the bounce language and use the asymptotic behavior of $\rho(\xi)$ and $\phi(\xi)$ for $\xi\rightarrow \infty$
(see {\it e.g.} appendix A of \cite{MPW}).

In the Minkowski case, remember that $-V_t'/D=\dot\rho$
asymptotes to $\dot\rho=1$ at $\xi\rightarrow \infty$, so that 
$
\lim_{\phi\rightarrow\phi_\pp}V_t'/D= -1.
$
Using that $V_t(\phi)$ approaches zero at 
$\phi_\pp $ 
as 
$V_t(\phi)\simeq 3m^2(\phi-\phi_\pp)^2/[2\log(\phi-\phi_\pp)]$ and $V_t'(\phi)\simeq 3m^2(\phi-\phi_\pp)/\log(\phi-\phi_\pp)$, one gets $1+V_t'/D\simeq (\kappa/4)(\phi-\phi_\pp)^2\log(\phi-\phi_\pp)$ so that $F_0(\phi_\pp)=0$. In conclusion, for Minkowski and AdS decays the vanishing of $F$ and $F_0$ at the
boundaries $\phi_0$ and $\phi_\pp$ proves that 
$S=\Delta S_E$, as promised.

Consider next the dS case, for which
\be
\Delta S_E=\int_{\phi_{0\pp}}^{\phi_0}{\it s}_E\,d\phi + \frac{24\pi^2}{\kappa^2V_\pp}\ ,
\ee
with 
\be
{\it s}_E=-108\pi^2\frac{(V-V_t)V}{D^3}\ ,
\ee
which is obtained translating (\ref{SE1}) to field space.

One also has
\bea
S&=&\int_{\phi_\pp}^{\phi_0}{\it s}\,d\phi \nonumber\\
&=&
\int_{\phi_{0\pp}}^{\phi_0}{\it s}\,d\phi+\frac{24\pi^2}{\kappa^2}\left(\frac{1}{V_\pp}-\frac{1}{V_{0\pp}}\right)\ ,
\eea
where ${\it s}$ is as given in (\ref{senew}) and
$V_{0\pp}\equiv V(\phi_{0\pp})$.

For this dS case one can check that
\be
{\it s} - {\it s}_E = \frac{dF_0}{d\phi}\ , 
\label{senewoldS}
\ee
with $F_0(\phi)$ given in (\ref{F0}). Integrating this in the interval $(\phi_{0\pp},\phi_0)$ one has
\be
 S - \Delta S_E  = F_0(\phi_0) - F_0(\phi_{0\pp}) - \frac{24\pi^2}{\kappa^2V_{0\pp}}\ .
\label{senewoldSint}
\ee
Now, at $\phi_0$ one still has $D=-V_t'$, so that $F_0(\phi_0)=0$,
while at $\phi_{0\pp}$ one has $V_t'>0$ so that $D=V_t'$, and $F_0(\phi_{0\pp})=-24\pi^2/(\kappa^2 V_{0\pp})$. Plugging this in 
(\ref{senewoldSint}) leads to the claimed equality $ S = \Delta S_E  $.

\bigskip
\paragraph{\bf $\bma{\S\,}$ A2. Thin-wall case\label{sec:tw}} 

When the potential difference between false and true vacua is very small, one expects to be in the thin-wall limit (with the bounce having a sharp transition between $\phi_0\simeq \phi_-$ and $\phi_\pp$ at some $\xi$). In such cases an analytical expression for the tunneling action can be obtained, in terms of the wall tension $\sigma$ \cite{CdL,Weinberg}.

The derivation of the analytic thin-wall expression for the tunneling action using the tunneling potential approach proceeds as follows.
Consider first the Minkowski and AdS cases.
When the barrier separating the false and true minima in $V$ is high compared to $V_t$ one has $V_t'\ll (V-V_t)'$. Using this, one can approximate the equation of motion for $V_t$, written in the form (\ref{EoMD}) as
\be
\frac{d}{d\phi}\log D \simeq \frac{(V-V_t)'}{2(V-V_t)}\ ,
\ee
which is readily integrated to get
\be
D^2 = (V_t')^2 + 6\kappa (V-V_t)V_t \simeq C (V-V_t)\ ,
\ee
where $C$ is an integration constant that can be expressed in terms of the wall-tension as follows.  Rewrite the previous equation as
\be
\sqrt{V-V_t}\simeq \frac{-V_t'}{\sqrt{C-6\kappa V_t}}\ .
\label{noV}
\ee
From this, one gets
\bea
\sigma &\equiv& \int_{\phi_\pp}^{\phi_0} \sqrt{2(V-V_t)}d\phi
=-\sqrt{2}\int_{\phi_\pp}^{\phi_0} 
\frac{V_t' d\phi}{\sqrt{C-6\kappa V_t}}\nonumber\\
&\simeq&
\frac{\sqrt{2}}{3\kappa}\left(\sqrt{C-6\kappa V_\mm}-\sqrt{C-6\kappa V_\pp}\right)\, ,
\eea
where $V(\phi_0)\simeq V_\mm$, 
valid in the thin-wall case, has been used.
From this, $C$ is extracted as
\be
C = 6\kappa V_\pp +\frac{1}{8\sigma^2}(4\Delta V -3\kappa \sigma^2)^2\ ,
\ee
where $\Delta V\equiv V_\pp-V_\mm$.

Using the results above, the tunneling action (\ref{action}) can be rewritten as
\be
 S_{ tw} = \frac{6\pi^2}{\kappa^2}\int_{\phi_\pp}^{\phi_\mm} d\phi \frac{\sqrt{V-V_t}}{V_t^2\sqrt{C}}\left(\sqrt{C}-\sqrt{C-6\kappa V_t}\right)^2\ .
\ee
This integral can be performed if one gets rid of the only $V$ appearance in the integrand using (\ref{noV})  to get
\be
 S_{ tw}  = \frac{12\pi^2}{\kappa^2 V_t}\left.\left(1-\sqrt{1-6\kappa V_t/C}\right)\right|_{V_t=V_\mm}^{V_t=V_\pp}\, .
\label{tw}
\ee
This agrees with the known result in the literature, see {\it e.g.} \cite{Weinberg,Fortin,Brown}. In the case of small gravitational effects an expansion in $\kappa$ gives
\be
 S_{ tw} = S_{ tw}^{(\kappa=0)}\left[1-\frac{3\kappa \langle V\rangle \sigma^2}{\Delta V^2}+{\cal O}(\kappa^2)\right]\, ,
\ee
where 
\be
S_{ tw}^{(\kappa=0)}=\frac{27\pi^2\sigma^4}{2\,\Delta V^3}\, ,
\ee
and $\langle V\rangle\equiv (V_\pp+V_\mm)/2$.

In the Minkowski limit ($V_\pp\rightarrow 0$) one recovers from the general expression (\ref{tw}) the Coleman result \cite{CdL}
\be
S_{tw}= \frac{S_{tw}^{(\kappa=0)}}{(1-\kappa/\kappa_c)^2}\ ,
\ee
with 
\be
\kappa_c\equiv \frac{4\Delta V}{3\sigma^2}\ ,
\ee
displaying the gravitational quenching of the decay for
$\kappa/\kappa_c\rightarrow 1$.
For $\kappa/\kappa_c>1$ the decay is forbidden. In the new formalism this critical value comes from the requirement that $D$ should be real. This translates into the inequality
\be
\frac{-V_t'}{\sqrt{-V_t}}>\sqrt{6\kappa (V-V_t)}\ ,
\ee
integration of which results in the condition $3\kappa \sigma^2< 4\Delta V$.

Generalizing this discussion to include also AdS decays, 
one gets  
\be
\kappa_c\equiv \frac{4\left(\sqrt{-V_\pp}-\sqrt{-V_\mm}\right)^2}{3\sigma^2}\ ,
\ee
with decays forbidden if $\kappa/\kappa_c>1$.

In the dS case the near degeneracy of false and true vacua does not necessarily imply the thin wall case. In terms of the tunneling potential approach, near-degenerate vacua do not necessarily imply that $V_t$ is very flat in this case. This is because, first, now there is no obstruction preventing $V_t$ to curve upwards and second,  the instanton part of the $V_t$ solution does not necessarily connect to $\phi_\pp$.  Nevertheless, when the instanton part of the dS tunneling dominates (for small enough $\kappa V_\pp$) with $\phi_{0\pp},\phi_t\rightarrow \phi_\pp$, the derivation of the thin-wall action proceeds as before.


\end{document}